\documentclass[twocolumn,amsmath,floatfix,prl,aps,superscriptaddress]{revtex4-1}
\usepackage{epsfig} 
\usepackage{amsmath}
\usepackage{mleftright}\mleftright
\usepackage{hyperref}
\usepackage{float}
\usepackage{times}
\usepackage[caption = false, labelformat = empty, farskip = -22pt]{subfig}
\usepackage{graphicx}
\usepackage{xcolor}

\begin{document} 

\title{ SNS Junctions along the BCS-BEC Crossover}

\author{Gautam Rai} 
\address{I. Institute of Theoretical Physics, University of Hamburg, 22761 Hamburg, Germany}

\author{Arman Babakhani } 
\address{Department of Physics and Astronomy, University of Southern California, Los Angeles, CA 90089-0484, USA}

\author{Ying Wang} 
\address{School of Physics and Electronic Engineering, Jiangsu University, Zhenjiang, 212013, PR China}

\author{ Stephan Haas} 
\address{Department of Physics and Astronomy, University of Southern California, Los Angeles, CA 90089-0484, USA}

\address{Constructor University, 
 Campus Ring 1, 28759 Bremen, Germany}

\author{Stefan Kettemann} 

\address{Constructor University, 
 Campus Ring 1, 28759 Bremen, Germany}

\date{\today}

\begin{abstract} 
We present a theory of SNS junctions, a normal metal sandwiched between two superconductors, along the crossover from the BCS to the BEC regime. We calculate the Josephson current as a function of the chemical potential relative to the band edge in the superconducting region, $\mu_S$, where the BEC phase is indicated by $\mu_S <0$. The chemical potential relative to the band edge in the normal metal, $\mu_N$, allows us to tune the junction between the SNS case ($\mu_N>0$) and the SIS case, where the superconductors are separated by a tunneling barrier. We find that there are Andreev levels in the BEC regime, as long as there is sufficient density of states in the normal region, i.e. when $\mu_N>\Delta$, where $\Delta$ is the amplitude of the superconducting order parameter. For 1D SNS junctions, we find the Josephson current $I_S$ carried by these Andreev levels to be a function of the ratio $\Delta/\Delta_d$, where $\Delta_d$ is the Andreev level spacing. At zero temperature, the Josephson current has a maximum on the BCS side of the transition where $\Delta$ is maximal. At finite temperature, however, we find that the maximum moves to the BEC side of the crossover. We identify the mechanism for this phenomenon to be the decrease in the number of Andreev levels at the BCS-BEC crossover, accompanied by an increase in excitation energy to the unoccupied levels, making it less likely that these states are thermally occupied. Thereby, at finite temperature, the Josephson current is more strongly reduced on the BCS side of the crossover, resulting in a maximal Josephson current at the BCS-BEC crossover.  
\end{abstract}

\maketitle


{\it Introduction.} 
Josephson junctions~\cite{Josephson1962} are an excellent platform to study the transport properties of weakly bound Cooper pairs in the BCS regime to tightly bound bosonic pairs in the BEC limit. 
The 
Josephson current 
through an SIS junction, two superconductors
 separated by a tunnel barrier, is known to exhibit a maximum at the BCS to BEC crossover~\cite{Miller2007,Weimer2015},
 in agreement with theoretical calculations~\cite{Spuntarelli2010,Zwerger2019}. SNS junctions,
 connecting a normal metal with two superconducting leads, 
 have not been considered 
 in the BEC regime so far, 
 even through intriguing questions arise regarding the fate of Andreev levels and
 the magnitude of the Josephson current in the SNS junction:
 How many Andreev levels are there in the SNS junction, and 
 is there a maximum in the Josephson current as the Cooper pairs are tuned into tightly bound bosons? 

 The general theory for the BCS-BEC crossover is well established. 
 Eagles~\cite{Eagles} solved BCS equations
 within a single-band semiconductor model and found a crossover to a BEC condensate as the doping concentration is lowered. The charge carriers form local pairs which condense into a BEC at low temperatures~\cite{Leggett1980}. Nozieres and Pistolesi~\cite{Nozieres} extended this theory to a two-band model, which was further extended for multi-band systems in Refs.~\cite{Hirschfeld2015,Chubukov2016,Loh2016,Efetov2017,Yerin2019}.
 Experimentally, the BCS-BEC crossover was 
 first realized in artificial atom systems 
~\cite{Regal2004,Zwierlein2004}.
 More recently, a 
 BCS-BEC crossover has also been experimentally observed 
 in thin films of Fe-based superconductors~\cite{Rinott2017}
 by chemical variation of the doping level and in layered nitrides by gate controlled doping~\cite{Nakagawa2018}. Furthermore, in magic-angle 
 twisted bilayer graphene superconductivity was discovered at low carrier concentrations, tunable by gate controlled doping~\cite{Cao2018}, which opens another venue to investigating the BCS-BEC crossover experimentally. 
 Recently, a novel 1D superconductor was discovered in twisted bilayer graphene~\cite{Barrier2024}.
 In Ref.~\cite{Lin2022}
 a spatial crossover
 between a BCS and BEC regime was observed by scanning tunneling microscopy in 
 FeSe thin films, deposited on a trilayer graphene substrate. 
 Diminished coherence peaks indicate a normal conduction region, sandwiched between regions in 
 a BCS phase (at a 
 temperature $T< T_C \approx 2K$) and 
 a BEC phase (at 
 $T< T^* \approx 1.4K$), forming an SNS junction.

 \begin{figure}
 \vspace{.0cm}
\includegraphics[width=0.5\textwidth]{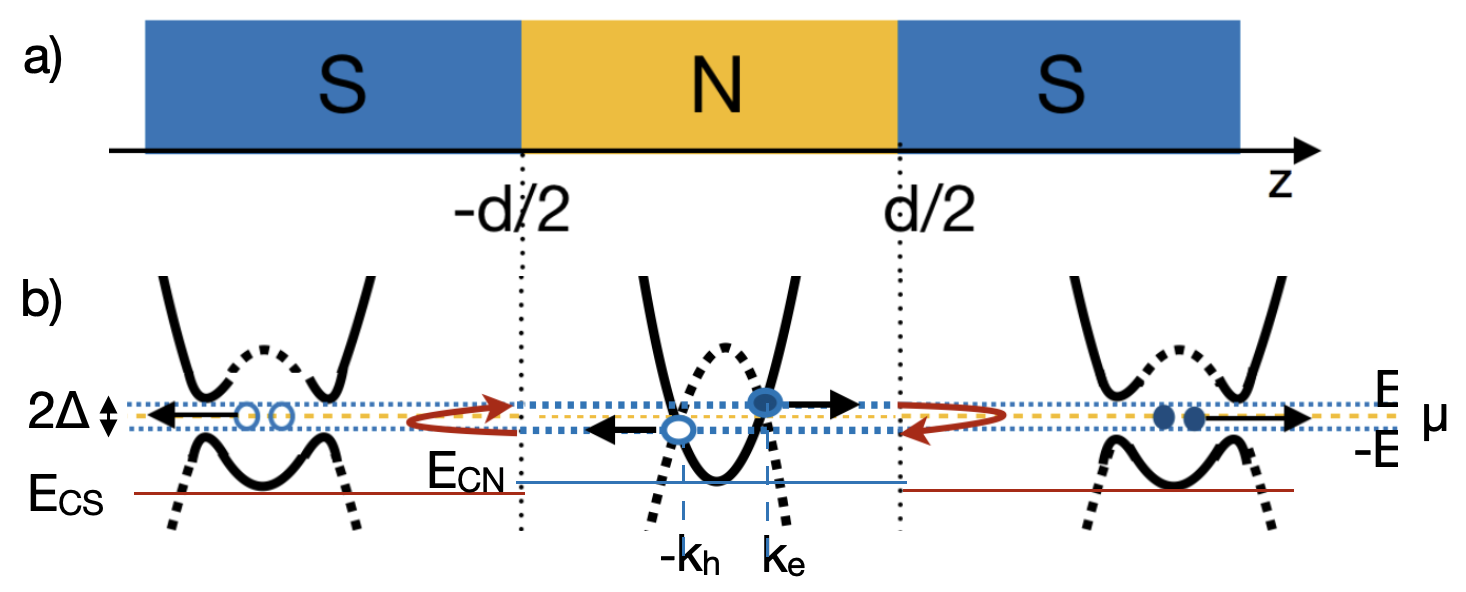}
 \vspace{-0.5cm}
\caption{ a) SNS-junction, with a normal region (N) sandwiched between two superconducting leads (S). b) electron and hole dispersions in each region. Blue lines: electron and hole energies relative to the chemical potential $\mu$, $E,-E$.
Red arrows: 
Andreev Scattering of electron with momentum $k_e$ to 
 hole with momentum $-k_h$ and back, forming a clockwise standing wave, an Andreev level. 
The anticlockwise standing wave, from $-k_e$ to $k_h$ and back, forms another Andreev level (not shown). Red, blue solid lines: band edge 
in the superconducting, normal region, 
$E_{CS},E_{CN}$,
respectively.}
\label{SNSAndreev}
 \vspace{-0.5cm}
\end{figure}

This motivates us to study SNS-junctions along the BCS-BEC crossover. We aim to 
derive their properties 
as a function of 
 the chemical potential relative to the band edge ($E_{CS}$) in the superconductors, $\mu_S=\mu - E_{CS}$, governing the BCS-BEC crossover. 
The crossover between SNS and SIS junctions, on the other hand, 
is governed by the 
 chemical potential $\mu_N=\mu - E_{CN}$,
 relative to the normal conductor band edge $E_{CN}$ 
 and the thickness of the normal region $d$, see Fig. \ref{SNSAndreev}.

{\it Bogoliubov-De Gennes Equations for SNS Junctions. }
 The BdG equations for two-component fermionic wave functions 
 are given by~\cite{Degennes1964,Spuntarelli2010}
\begin{equation}
\left( 
\begin{array}{cc}
\mathcal{H}(\mathbf{r}) & \Delta(\mathbf{r}) \\
\Delta(\mathbf{r})^{*} & - \mathcal{H}(\mathbf{r}) 
\end{array} 
\right)
\left( \begin{array}{c}
u_{n}(\mathbf{r}) \\
v_{n}(\mathbf{r}) 
\end{array} 
\right) 
= E_{n}
\left( \begin{array}{c}
u_{n}(\mathbf{r}) \\
v_{n}(\mathbf{r}) 
\end{array} 
\right) . 
\label{B-dG-equations} 
\end{equation}
Here, $\mathcal{H}(\mathbf{r}) = - \nabla^{2}/(2m) + V(\mathbf{r}) - \mu$, with fermion mass
$m$ and external potential $V(\mathbf{r})$. 
 $v_{n}$
 is the probability amplitude that there is a Cooper pair 
 in the state indexed by $n$, and $u_{n}$ the amplitude 
 that there is no Cooper pair in that state. The functions $\{u_{n}(\mathbf{r}),v_{n}(\mathbf{r})\}$ obey orthonormality conditions.
The local order parameter $\Delta(\mathbf{r})$ is determined via the \emph{self-consistency condition}~\cite{schrieffer} which at $T=0$~K is given by 
\begin{equation}
\Delta(\mathbf{r}) = U \sum_{n} u_{n}(\mathbf{r}) 
v_{n}(\mathbf{r})^{*}, n(\mathbf{r}) = 
2 \sum_{n} \, |v_{n}(\mathbf{r})|^{2} \label{self-consistency},
\end{equation}
where $U>0$ is the local fermionic attraction strength. The chemical potential $\mu$ is determined by the conservation of particle number $N$, 
 as obtained by 
integrating the number density $n({\bf r})$ in Eq. (\ref{self-consistency})
over 
the sample volume,
$N = \int d {\bf r} n ({\bf r})$.
 In Fig. \ref{SNSAndreev} b), 
 we sketch the
 local energy dispersions of electrons (solid line)
and holes (dashed lines) in
 the SNS junction.
The potential difference between the 
 leads (that become superconducting once the pairing interaction is turned on) and the normal region is given by $V=E_{CS}-E_{CN}$ (see 
 Fig. \ref{SNSAndreev} b)).
 We distinguish the doping level parameter $\epsilon$, the Fermi energy before the attractive interaction is turned on~\cite{Nozieres},
 and $\mu$, 
 the self consistently calculated chemical potential. 
 $\mu$ is lowered by the binding energy of the Cooper pairs and is a nonlinear function of the doping level $\epsilon$~\cite{Nozieres,Niroula2020}.
 We denote $\epsilon_S = \epsilon -E_{CS}$, 
and $\epsilon_N = \epsilon -E_{CN}$, as the doping levels relative to the conduction band edges in the leads and in the normal region, respectively.
The chemical potential
$\mu$ 
in the superconducting region is 
shifted down relative to 
$E_{CS}$,
lowering 
$\mu_S = \mu - E_{CS}$
due to the pairing energy in the superconducting condensate.
 As the doping level $\epsilon$ 
 is lowered, one enters 
 the BEC phase in the leads, 
 when the chemical potential 
drops below the conduction band in the superconductor, that is, when $\mu_S <0$.
The chemical potential relative to the 
band edge in the normal region
 $\mu_N$
 remains constant to ensure charge neutrality.
Accordingly, the potential difference between the superconducting and normal region 
becomes larger $V_{\rm eff} >V$,
due to the pairing energy in the superconducting condensate.

For a bulk superconductor, the quasiparticle energy is
given by $E_{\bf k} = \sqrt{\xi_{\bf k}^2 + \Delta^2} ,$
with 
$\xi_{\bf k} = {\bf k}^2/(2m)- \mu_S.$
The density of states of quasiparticle excitations 
 in the BCS regime 
 has a gap $2 \Delta$ 
 with two sharp coherence peaks, as sketched in Figs. 
\ref{figDOSSNS} a) and b). In the BEC phase,
$\mu_S < 0$, the energy gap 
is enlarged to 
$\Delta_G = \sqrt{\mu_S^2+ \Delta^2}> \Delta.$
 There, the coherence peaks in the density of states are reduced, and especially the lower one is strongly diminished (see Figs. 
\ref{figDOSSNS} c) and d))
~\cite{Lin2022}.
 \begin{figure}[h]
\includegraphics[width=4.25cm]{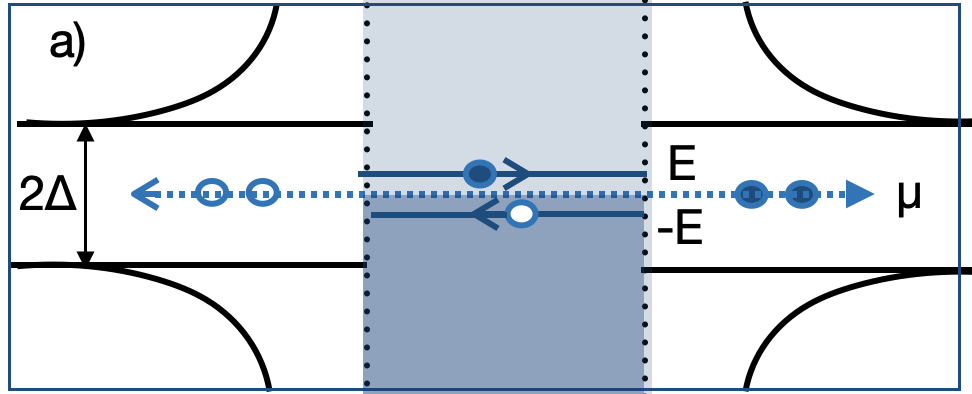}
\includegraphics[width=4.25cm]{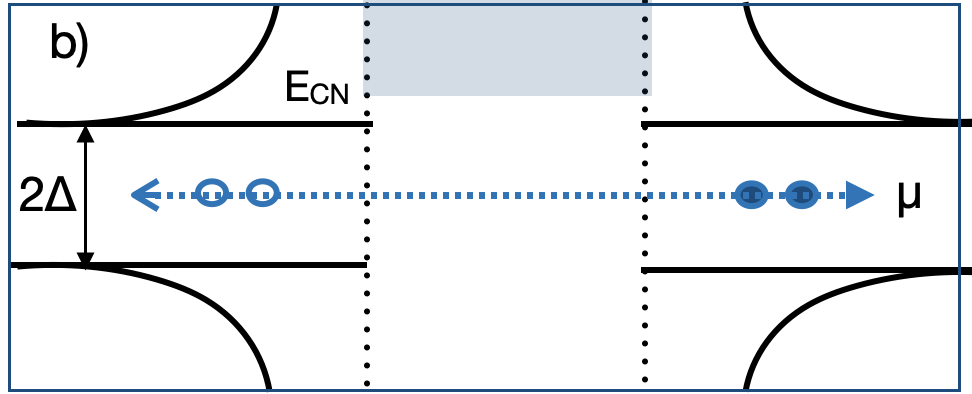}
\caption{Schematic density of states of 
a) BCS-N-BCS junction, b) BCS-I-BCS junction with tunneling barrier $E_{CN}-\mu$.
}
 \label{figDOSSNS} 
\end{figure}

 \begin{figure}[h]
\includegraphics[width=4.25cm]{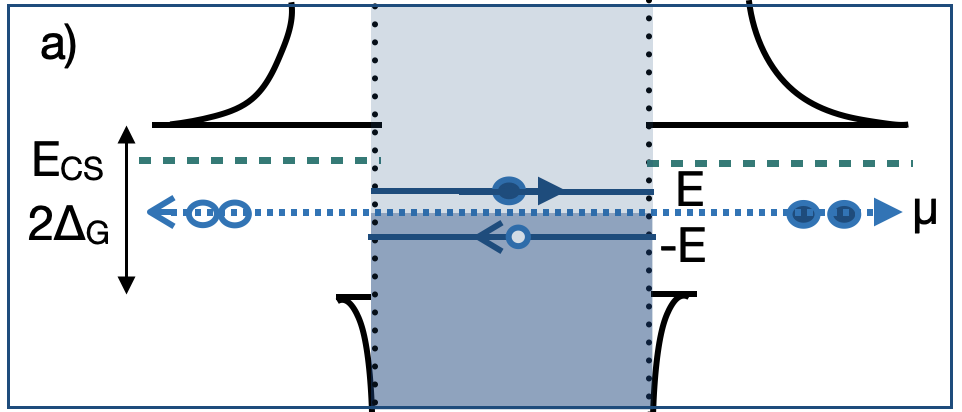}
\includegraphics[width=4.25cm]{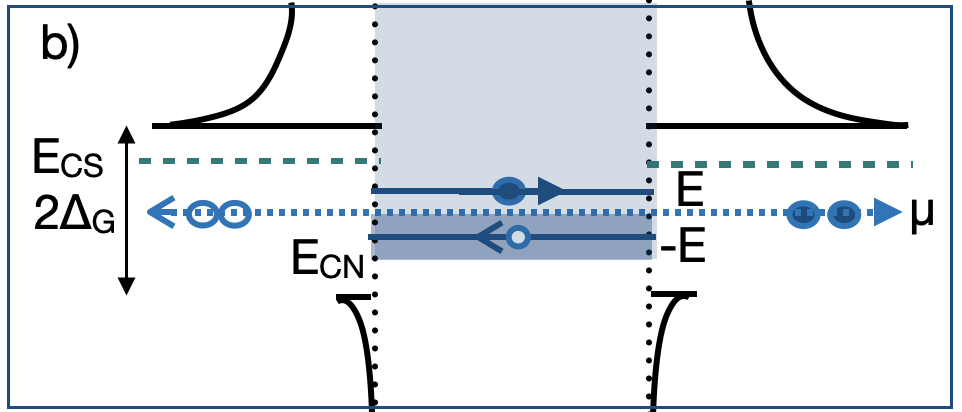}
\caption{Schematic density of states of 
a) BEC-N-BEC junction with $\mu_N = \mu-E_{CN} > \Delta_G$, b) 
BEC-N-BEC junction with $\mu_N = \mu-E_{CN} < \Delta_G.$
}
 \label{figDOSSNS} 
 \vspace{-.5cm}
\end{figure}
 
The solution 
of Eq. (\ref{B-dG-equations}) for an
SNS junction
is known in the BCS regime:
 an electron in the normal region with energy $E< \Delta$ forms a standing wave
with a hole, with energy $-E$, 
being reflected at the 
 interface by Andreev scattering, converting the electron into the hole and vice versa, as indicated by the red arrows in Fig. \ref{SNSAndreev} b). 
The electron and hole amplitudes are for $ \{z>d/2,z<-d/2\},$ respectively, given by 
$
\left( \begin{array}{c}
\psi_{e}(z) \\
\psi_{h}(z)
\end{array} 
\right) = \{ c_+ \exp(i k_+ (z-d/2)) \left( \begin{array}{c}
 e^{i \phi_R} \\
e^{-i \varphi_E} 
\end{array} 
\right), d_+ \exp(i k_- (z+d/2))\left( \begin{array}{c}
e^{-i \varphi_E} \\
e^{-i \phi_L} 
\end{array} 
\right) \}, \label{right2} 
$
with 
$\varphi_E = \arccos(E/\Delta),$, $k_+ = (2 m (\mu+i \sqrt{\Delta^2-E^2}))^{1/2}$, $Im k_+ >0$ and 
 $k_- = (2 m (\mu-i \sqrt{\Delta^2-E^2}))^{1/2}$, $Im k_- <0$.
 The electron and hole energies are sketched in Fig. \ref{SNSAndreev} b) for the example of a single Andreev bound state in the normal region.
The electron moving to the right and the hole moving to the left 
 have equal amplitudes, but different phases $\phi_L,\phi_R$. 
Next, we match the amplitudes at $z= \pm d/2$ with superpositions of plane waves in the normal region
 with 
 electron momenta $k_e = (2 m (E+ \mu_N))^{1/2}$
and hole momenta $k_h = (2 m (\mu_N-E))^{1/2}$.
We thus obtain the condition for the Andreev level energy, expanding $\Delta k = k_e-k_h$ to first order in $E/\mu_N$ 
for $E\ll \mu_N,$
~\cite{Andreev1966,Kulik1969,Sauls2018}
\begin{equation} \label{enplus}
 \frac{E^{+/-}_n}{\Delta} = 
 \frac{\Delta_d}{\Delta}
 \left( n + \frac{1}{\pi}\arccos (\frac{E^+_n}{\Delta}) \pm \frac{\Delta \phi}{2 \pi} \right), 
\end{equation}
with the phase difference $\Delta \phi = \phi_R-\phi_L$. 
Andreev levels close to the 
center of the gap
(where $|E| \ll \Delta$, $\arccos (E/\Delta) \approx \pi/2$) 
have spacing 
$\Delta_d = \pi \hbar v_{FN}/d$
with $v_{FN} = \sqrt{2 \mu_N/m}$
 the Fermi velocity in the normal region. 
$\Delta_d$
is twice the energy level spacing in a 1D
normal conductor of length $d$. 
The sign $\pm$ corresponds to standing waves with clockwise (+) 
and anticlockwise (-) Andreev scattering. 
 \begin{figure}[h]
\includegraphics[width=9cm]{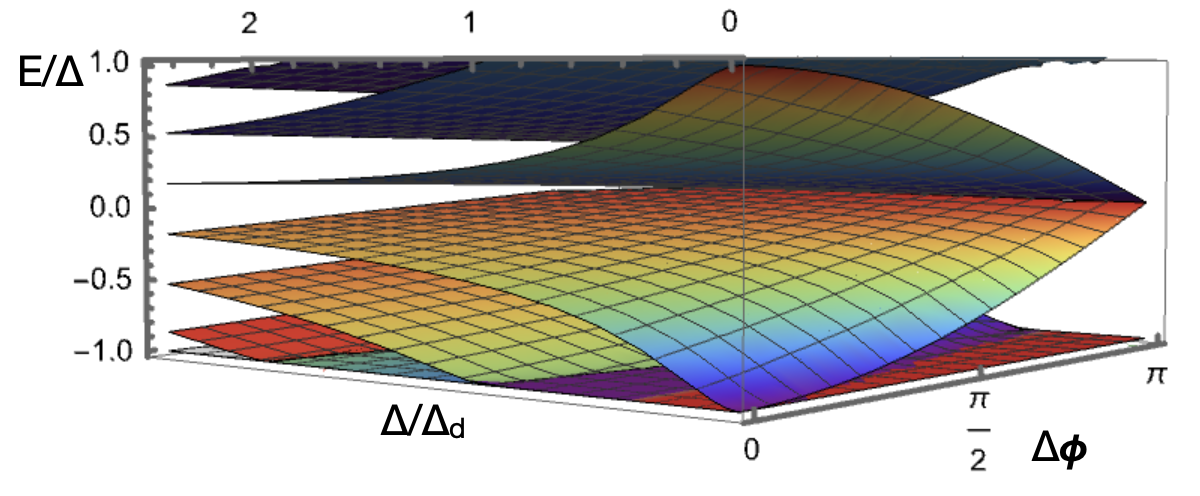}
\caption{
 Andreev levels, Eq. (\ref{enplus}),
 as function of 
 phase difference $\Delta \phi$
 and 
 ratio of order parameter and level spacing 
 $\Delta/\Delta_d= d/(\xi \pi)$.
 }
 \label{figALSNS} 
\end{figure}
In Fig. \ref{figALSNS},
 Andreev levels, solutions of Eq. (\ref{enplus}),
are plotted
as a function of the
 phase difference $\Delta \phi$
 and the ratio
 $\Delta/\Delta_d =d/(\xi \pi)$, 
 proportional to the ratio between length $d$
 of the normal metal and 
 the coherence length $\xi = \hbar v_{\rm F}/\Delta = \pi \xi_0,$
 where $\xi_0$
 is the Pippard coherence length in the BCS regime. 

In the BEC regime, $\mu_S <0,$ we
 find that 
 Andreev levels still exist 
 as solutions of Eq. (\ref{enplus}) for $|E| < \Delta <\Delta_G$,
 as long as there is sufficient density of states in the normal region, that is $\mu_N > \Delta_d$
 and $\Delta_d < \Delta.$
For larger energies $E$,
when $\Delta < |E| < \Delta_{\rm G},$
 Andreev and normal scattering mixes.
Then, 
 electron and hole amplitudes are
 no longer equal, so that these energy levels depend only weakly on the phase difference $\Delta \phi.$ 

 Both in the BCS and BEC regime, the number of Andreev levels is given by 
 $N_A = 2 Int[2 \Delta/\Delta_d]$ for 
 $\mu_N > \Delta$.
 The factor $2$ accounts for the pairs of Andreev levels, corresponding to the clockwise and counterclockwise standing waves, Fig. \ref{SNSAndreev}. 
Andreev levels can only form 
when 
 there is sufficient density of states around the chemical potential in the normal region.
 This is the case 
 both in the BCS limit, as shown in Fig. \ref{figDOSSNS} a) and in the BEC limit Fig. \ref{figDOSSNS} c), as long as $\mu_N > \Delta.$
However, when $\mu_N$ is so close to the band edge that $0<\mu_N<\Delta$,
electrons of energy $E$ can only find holes of energy $-E$ in a band of width $2 \mu_N< 2 \Delta$ around the chemical potential. Then, 
the number of
 Andreev states reduces to 
 $N_A = 2 Int[2 \mu_N/\Delta_d]$. 
 Inserting $\Delta_d = \hbar \pi v_{\rm FN}/d$,
we find
\begin{equation}
N_A = 2
\begin{cases} 
 Int[\frac{d}{\pi a_0} \frac{\Delta}{\sqrt{\mu_N t}}] ~{\rm for}~
 \Delta < \mu_N ,\\
Int[ \frac{d}{\pi a_0}
\sqrt{\frac{\mu_N}{t}}] ~ {\rm for}~
 0<\mu_N< \Delta , 
 \end{cases} 
 \label{na}
\end{equation}
where $t= \hbar^2/(2 m a_0^2)$,
and $a_0$ is the lattice spacing.
Note that for a larger number of Andreev levels, 
their 
 dependence on the phase difference $\Delta \phi$ becomes weaker, as is observed in Fig. \ref{figALSNS}.
For short junctions, when 
$\Delta_d > \Delta,$ 
Andreev levels cannot form, in which case the junction is in the SIS limit.

{\it
Josephson Current. }
In the BCS regime, the Josephson current through 
 an SIS junction is given by~\cite{Josephson1962,Sauls2018}
 \begin{equation} \label{IsJ}
 I_s = G_T \pi \frac{\Delta}{e } \sin \Delta \phi,
 \end{equation}
 with the tunneling conductance 
 $G_T = G_0 \exp (- 2 d (2m(V_0-\mu))^{1/2})$, where $G_0 = 2 e^2/h.$
 It decays exponentially with tunneling barrier width $d$ and height $V_0-\mu$. 
 In 
the BCS-BEC crossover regime, 
the Josephson current 
through SIS junctions 
is 
known to be maximal at the crossover point~\cite{Zwerger2019,Setiawan2022,Lewandowski2023,Spuntarelli2010,Pascucci2020}, as confirmed 
 experimentally in cold atom systems~\cite{Miller2007,Weimer2015}.
 This is explained by the fact that 
 the current is limited by the available excitations in the junction. Quasiparticle excitations give an upper limit, 
 which increases 
 when moving from BCS to BEC, 
while sound wave excitations give a decreasing upper limit when moving from BCS to BEC.
This results in a maximal Josephson current at the BCS-BEC crossover~\cite{Spuntarelli2010}.

The Josephson current in an SNS junction with a Sharvin contact 
is given by~\cite{Sauls2018} 
\begin{equation} \label{sauls}
I_s = G_N \pi \sum_{n} 
f(E_{n}) \frac{d E_{n}/e}{d \Delta \phi},
\end{equation}
 where
 $G_N$ is the normal state conductance of the
 junction.
A Sharvin contact has a ballistic channel with conductance
$G_N=2 e^2/h$.
 Thus, the Josephson current is carried by occupied
 Andreev levels
 $E_{n}( \Delta \phi),$ Eq. (\ref{enplus}),
 which each contribute a term proportional to
 their slope
 in the phase
 difference $\Delta \phi$.
 We see in Fig. \ref{figALSNS}
 that subsequent Andreev levels have slopes of opposite sign, 
 so that the current is expected to decrease 
 with the number of occupied Andreev levels. 
Eq. (\ref{sauls}) implies that the Josephson current is mostly carried by occupied Andreev levels, with a small contribution carried by the continuum states located in the energy range outside of the superconductor energy gap, which depends only weakly on $\Delta \phi$. For a more general formulation 
see Refs.~\cite{Loefwander2001,Heikkilael2002,Nikolic2019}. 

 \begin{figure}[h]
\includegraphics[]{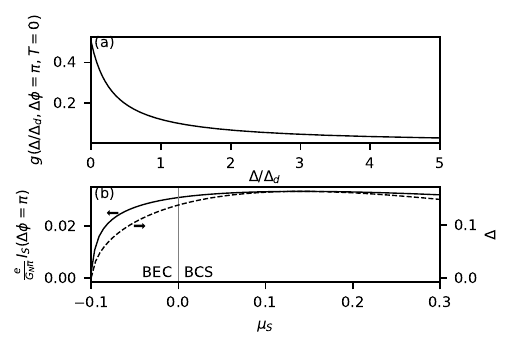}
 \vspace{-0.5cm}
\caption{
a) The universal function $g(\Delta/\Delta_d,\Delta \phi =\pi,T=0)$. 
b) Order parameter $\Delta$ 
of a 1D superconductor as function of $\mu_S$
 in units of $t$, and Josephson current $I_s$ through an SNS junction 
 between 1D superconductors 
at $T=0K$ at $\Delta \phi = \pi$
as function of $\mu_S$
 in units of $t$.
We set
 $U= 4 \pi t/10,$ $\mu_N= 0.5t$ and 
 $\Delta_d/t =1/(2\sqrt{2}).$
}
 \vspace{-0.cm}
 \label{figISSNS} 
\end{figure}

 According to 
 Eq. (\ref{sauls}) 
 the Josephson current 
is limited by the conductance of the normal metal multiplied by an effective ``voltage", 
 $\Delta/e$,
giving the inequality
 $|I_s| \le G_N \Delta/e.$
Therefore, it is convenient to rewrite 
 Eq. (\ref{sauls}) 
 as 
 \begin{equation} \label{scaling}
 I_s = G_N \pi \frac{\Delta}{e } 
 g(\Delta/\Delta_d, \Delta \phi,T).
 \end{equation}
 $g(\Delta/\Delta_d, \Delta \phi,T)$,
 is according to Eq. (\ref{enplus}) 
 a universal function of the ratio 
 $\Delta/\Delta_d$ 
 for given 
 $\Delta \phi$ and temperature $T.$
 Thus, the Josephson current depends on 
 the junction specific parameters $\Delta,\mu_N,d$
 and the density of states only through the ratio $\Delta/\Delta_d,$ which also determines the number of 
Andreev levels $N_A,$ Eq. (\ref{na}).
In Fig
 \ref{figISSNS} a) this function is seen for $\Delta \phi = \pi$ and $T=0K$
to be a monotonously decaying function of $\Delta/\Delta_d.$

 We thus identified two 
opposing mechanisms on the magnitude of the Josephson current in the BCS-BEC crossover regime: 
The Josephson current is proportional to 
1) the order parameter $\Delta$ 
and 2) a universal function $g(\Delta/\Delta_d, \Delta \phi,T)$,
which 
decreases with the ratio 
 $\Delta/\Delta_d$.

Let us consider as an example
an SNS junction between 1D superconductors.
The density of states of 
 a 1D wire is
$
 \rho(\epsilon_k) = a_0 (m/(2 \epsilon_k))^{1/2}/\pi,$
 for $\epsilon_k>0.$
Inserting that into Eq. (\ref{self-consistency})),
we find the order parameter as function of $\mu_S$, shown 
 in Fig. \ref{figISSNS} b), with a maximum $\Delta$ 
 on the BCS side, $\mu_S>0$. This maximum is due to the increase of density of states at the band edge in this 1D model, followed by 
 the decrease of $\Delta$ when $\mu_S$ is tuned into the BEC regime, $\mu_S<0.$
Calculating the contribution of Andreev levels to the 
Josephson current, Eq. (\ref{sauls}), we find that 
at $T=0K$ it is a saw tooth function, with maximal 
amplitude at $\Delta \phi =\pi,$ where it jumps to the value with opposite sign. 
In Fig. \ref{figISSNS} b),
the amplitude at $\Delta \phi =\pi$
is shown 
 as function of $\mu_S$ in units of $t$. 
We find a maximal Josephson current amplitude, 
 at a value of $\mu_S$ close to the one where the order parameter is maximal.

 \begin{figure}[h]
\includegraphics[width=8cm]{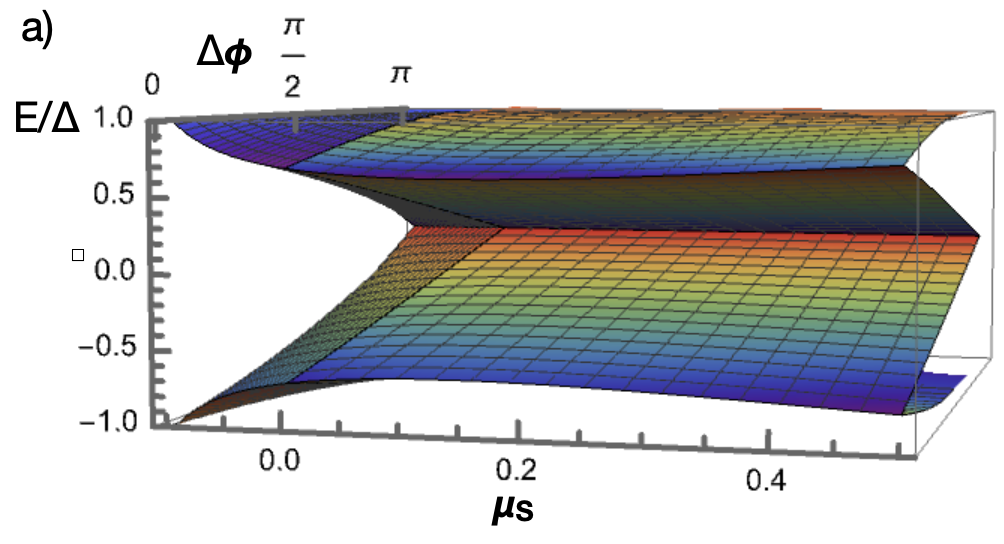}
\includegraphics[width=8cm]{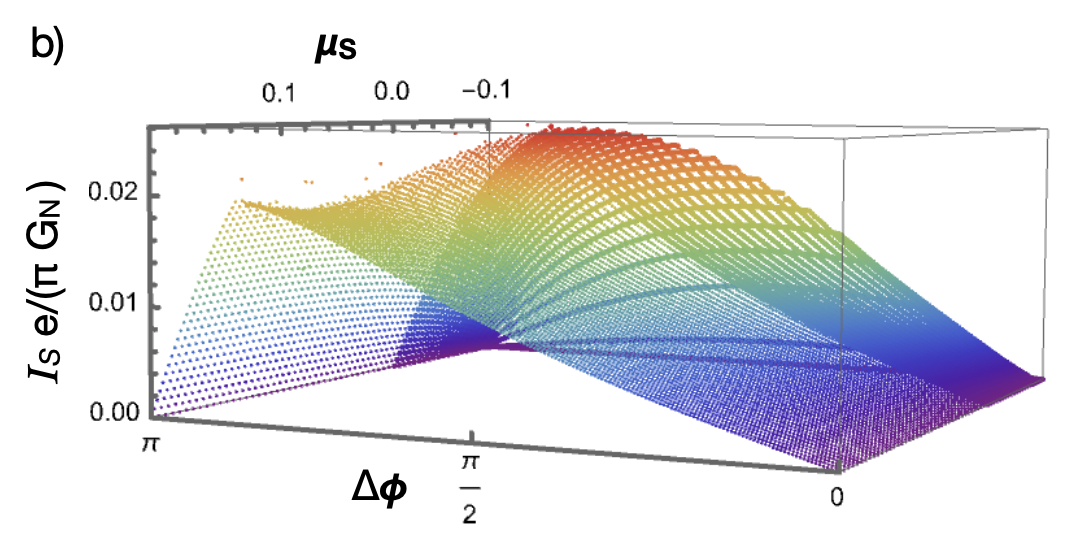}
\includegraphics{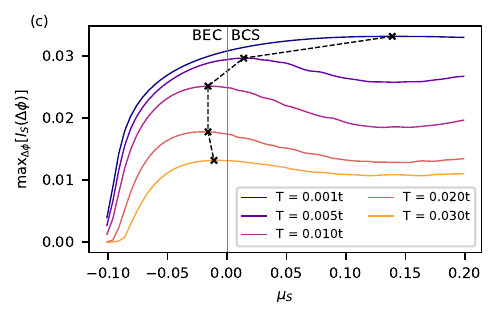}
 \vspace{-0.5cm}
\caption{a)
Andreev levels, energy 
 in units of $\Delta,$ 
as function of phase difference $\Delta \phi$ and chemical potential $\mu_S$. 
b)
Josephson current $I_s$ Eq. (\ref{sauls})
in 1D SNS-junctions
 for temperature $T = 0.01t$
as function of phase difference $\Delta \phi$ and chemical potential 
$\mu_S$. 
c) The maximal (with respect to $\Delta \phi$) Josephson current 
as function of $\mu_S$ for different 
 temperatures $T$. The crosses mark the maximum along the BCS-BEC crossover (with respect to $\mu_S$).
 We set
 $U= 4 \pi t/10,$ $\mu_N= 0.5t$ and 
 $\Delta_d/t =1/(2\sqrt{2}).$
} \vspace{-.5cm}
 \label{figIST} 
\end{figure}

To calculate the 
Josephson current at finite temperature $T>0K$
along the BCS-BEC crossover we insert the Fermi function
 $f(E_n,T) = 1/(1+ \exp(E_n(\Delta \phi)/T)) $
 in Eq. 
 (\ref{sauls}), with temperature $T$ in units of $t$. To take into account the temperature dependence of the order parameter, we insert 
 $\Delta(T) = \Delta(0) (1-1.76 ~ T/\Delta(0))^{1/2},$
 where we used the relation to the mean field critical temperature $T_c = c \Delta(0),$ where $c=1/1.76$~\cite{schrieffer}. 
 We see a decrease with increasing temperature. The jump 
 at $\Delta \phi =\pi$ becomes smooth. 
In Fig. \ref{figIST} a), we plot the energy of 
Andreev levels
 in units of $\Delta,$ 
as function of phase difference $\Delta \phi$ and chemical potential $\mu_S$. 
 As $\mu_S$ moves into the BEC regime, the number
 of Andreev levels reduces and their dependance on $\Delta \phi$ becomes thereby stronger. 
 At finite temperature, that effects the magnitude of the Josephson effect in another way: the excitation energy to unoccupied Andreev levels increases as $\mu_S$ is lowered. Therefore we expect that in the BCS regime, the Josephson current is more diminished with increasing temperature, since the excitation energy to unoccupied Andreev levels, which contribute to the current with an opposite sign, is lower there than in the BEC regime
 $\mu_S<0.$
Indeed
in Fig. \ref{figIST} b) 
we see that 
 the maximal Josephson current is located on the BEC side of the crossover, where we plot the 
Josephson current $I_s$ 
 for temperatures $T = 0.01t$
as function of phase difference $\Delta \phi$ and chemical potential 
$\mu_s$ in units of $t$.
In Fig. \ref{figIST} c) 
the maximal (with respect to $\Delta \phi$) Josephson current 
is plotted 
as function of $\mu_S$ for different 
 temperatures $T$. 
 We find that the maximum moves to the BEC side of the BCS-BEC crossover as the temperature is increased.
In the Supplementary Material, 
 we show results for a system with a band of constant density of states, where we also find a maximal Josepshon current,
 although it remains on the BCS side, 
 close to the BCS-BEC crossover. 

 {\it Conclusions and Discussion.}
We have analyzed SNS-junctions along the BCS-BEC crossover regime. We showed that 
 Andreev levels exist as long as there is sufficient density of states in the normal metal. 
Furthermore, we find 
a maximum in the Josephson current which moves 
 with increasing temperature into the BEC regime. 
The mechanism for this phenomenon is
identified as the decrease of the number of Andreev levels at the BCS-BEC crossover with an accompanying increase in excitation energy, thus making thermal excitations less likely on the BEC side. In a forthcoming publication, we will report the results of a numerical analysis of the BdG equations for SNS junctions on a tight-binding model~\cite{Babakhani}.
 
In Josephson junctions
with more than one transverse channels, 
there is finite angle Andreev scattering, 
which can be described by semiclassical transport theory, solving the Eilenberger equations
~\cite{Nikolic2019}. This method also allows to include finite transparency of the junction and disorder scattering which is present in any real sample and may reduce the Josephson current further. In future work, we plan to extend this theory and apply it to the BCS-BEC crossover regime. 
Furthermore, 1D superconductors are 
experiencing 
quantum and thermal fluctuations,
which requires to go beyond the BdG equations, as reviewed in Ref. 
~\cite{Aru2008}. To study its effect on the Josephson current, one needs to extend the theory of SNS junctions including the quantum fluctuations in the BCS-BEC crossover regime, which will be the
 subject of further investigations. 

\acknowledgments

S.K. gratefully acknowledges support from DFG KE-807/22-1. This work was supported by the US Department of Energy under grant number DE-FG03-01ER45908. 

\end{document}